%% file: Tsurumi_PRL_main.tex
\documentclass[aps, prl, dvipdfmx, superscriptaddress, preprint]{revtex4-1}
\usepackage{color}
\include{settings}

\begin{document}
	\title{Terahertz Raman spectroscopy probe of intermolecular vibration in high-mobility organic semiconductors under uniaxial strain}
	\author{Junto Tsurumi}
	\email{TSURUMI.Junto@nims.go.jp} 
	\affiliation{International Center of Materials Nanoarchitectonics (MANA), National Institute for Materials Science (NIMS), 1-1 Namiki, Tsukuba 305-0044, Japan}
	\affiliation{Material Innovation Research Center (MIRC) and Department of Advanced Materials Science, Graduate School of Frontier Sciences, The University of Tokyo, 5-1-5 Kashiwanoha, Kashiwa, Chiba 277-8561, Japan}

	\author{Takayoshi Kubo}
	\affiliation{Material Innovation Research Center (MIRC) and Department of Advanced Materials Science, Graduate School of Frontier Sciences, The University of Tokyo, 5-1-5 Kashiwanoha, Kashiwa, Chiba 277-8561, Japan}
	\affiliation{National Institute of Advanced Industrial Science and Technology (AIST)-The University of Tokyo Advanced Operando-Measurement Technology Open Innovation Laboratory (OPERANDO-OIL), AIST, 5-1-5 Kashiwanoha, Kashiwa, Chiba 277-8561, Japan}
	
	\author{Hiroyuki Ishii}
	\affiliation{Institute of Applied Physics and Tsukuba Research Center for Interdisciplinary Materials Science, University of Tsukuba, 1-1-1 Tennoudai, Tsukuba, Ibaraki 305-8573, Japan}
	
	\author{Masato Mitani}
	\affiliation{Material Innovation Research Center (MIRC) and Department of Advanced Materials Science, Graduate School of Frontier Sciences, The University of Tokyo, 5-1-5 Kashiwanoha, Kashiwa, Chiba 277-8561, Japan}
	
		\author{Toshihiro Okamoto}
	\affiliation{Material Innovation Research Center (MIRC) and Department of Advanced Materials Science, Graduate School of Frontier Sciences, The University of Tokyo, 5-1-5 Kashiwanoha, Kashiwa, Chiba 277-8561, Japan}
	\affiliation{National Institute of Advanced Industrial Science and Technology (AIST)-The University of Tokyo Advanced Operando-Measurement Technology Open Innovation Laboratory (OPERANDO-OIL), AIST, 5-1-5 Kashiwanoha, Kashiwa, Chiba 277-8561, Japan}
	\affiliation{JST, PRESTO, 4-1-8 Honcho, Kawaguchi, Saitama 332-0012, Japan}
	
	\author{Shun Watanabe}
	\email{swatanabe@edu.k.u-tokyo.ac.jp}
	\affiliation{Material Innovation Research Center (MIRC) and Department of Advanced Materials Science, Graduate School of Frontier Sciences, The University of Tokyo, 5-1-5 Kashiwanoha, Kashiwa, Chiba 277-8561, Japan}
	\affiliation{National Institute of Advanced Industrial Science and Technology (AIST)-The University of Tokyo Advanced Operando-Measurement Technology Open Innovation Laboratory (OPERANDO-OIL), AIST, 5-1-5 Kashiwanoha, Kashiwa, Chiba 277-8561, Japan}
	\affiliation{JST, PRESTO, 4-1-8 Honcho, Kawaguchi, Saitama 332-0012, Japan}
	
	\author{Jun Takeya$^*$}
	\email{takeya@k.u-tokyo.ac.jp}
	\affiliation{International Center of Materials Nanoarchitectonics (MANA), National Institute for Materials Science (NIMS), 1-1 Namiki, Tsukuba 305-0044, Japan}
	\affiliation{Material Innovation Research Center (MIRC) and Department of Advanced Materials Science, Graduate School of Frontier Sciences, The University of Tokyo, 5-1-5 Kashiwanoha, Kashiwa, Chiba 277-8561, Japan}
	\affiliation{National Institute of Advanced Industrial Science and Technology (AIST)-The University of Tokyo Advanced Operando-Measurement Technology Open Innovation Laboratory (OPERANDO-OIL), AIST, 5-1-5 Kashiwanoha, Kashiwa, Chiba 277-8561, Japan}

\begin{abstract}
Terahertz Raman spectroscopy was performed on high-mobility organic single-crystal semiconductors, by which the phonon energy at the $\Gamma$ point was qualified as a function of external uniaxial strain.
The observation of peak shifts in Raman modes under uniaxial strain revealed that application of an external strain can effectively tune the intermolecular vibration that particularly correlates electron--phonon coupling. 
Terahertz Raman spectroscopy conducted in conjunction with molecular dynamics simulation provides an in-depth understanding of the recently discovered very large strain effect on charge conduction in high-mobility organic semiconductors.
\end{abstract}
\maketitle

Small molecular organic semiconductors (OSCs) can form a single crystal that is a highly symmetric assembly of $\pi$-conjugated molecules bound by the weak van der Waals interaction. 
Carriers, particularly in single-crystal OSCs, undergo band transport via a slight overlap of $\pi$ orbitals between neighboring molecules. 
Transfer integral $t$, which determines the transfer probability between molecules and is consistent with the effective mass in band theory, is estimated to be $\sim$\SI[number-unit-product]{100}{meV} at most in any existing small-molecule OSC. 
This considerably small value of $t$ in OSCs, which is approximately 1/10 of that in inorganic semiconductors~\cite{vogl1983semi}, changes moment by moment because of the thermal vibration of molecules, and the thermal fluctuation of the transfer integral, $\Delta t/t$, is considered to be one of the key factors determining transport properties~\cite{troisi2006charge, ciuchi2011transient, ishii2012wave, fratini2016transient}.
It has been concluded so far that transiently localized carriers ranging in size from a few to dozens of molecules can limit mobility, allowing low-frequency intermolecular vibration to effectively modulate transfer integral in OSCs~\cite{ciuchi2011transient,fratini2016transient}.

The thermal fluctuation of the transfer integral, $\Delta t/t$, for a typical OSC, pentacene, is estimated to be $20$\% at room temperature~\cite{sakai2016emergence}, which is frequent enough to break the coherence of the wave function of the carriers.
As a result, carriers in pentacene, even in its single-crystal form, undergo hopping transport rather than band transport, which has been verified by the non-ideal Hall effect signals~\cite{sakai2016emergence}. 
In contrast, carrier transport in pentacene under the application of high pressure at low temperature approaches coherent, bandlike transport.
Given this fact, it is reasonable to suppose that the effective suppression of the thermal fluctuation of the transfer integral, $\Delta t/t$, plays a vital role in carrier coherence.

\begin{figure}[htbp]
	\centering
	\includegraphics[pagebox=artbox]{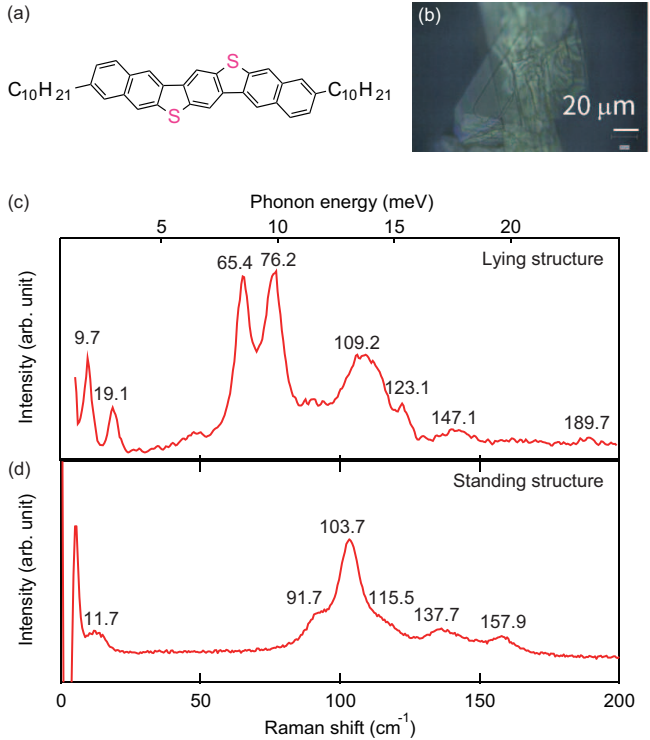}
	\caption{(a) Molecular structure of \tm{26}. (b) Optical microscopy image of the \tm{26}\ bulk crystal. (c) Terahertz Raman spectra of the two polymorphs of \tm{26}: ``lying'' and ``standing'' structures.}\label{fig:1}
\end{figure}

The importance of the thermal fluctuation of the transfer integral is being increasingly recognized in synthetic chemistry. For example, the target material in this paper, \tmfull{26} (\tm{26}) (Fig.~\ref{fig:1})~\cite{mitsui2014high}, is designed to reduce intermolecular vibrations by the introduction of an N-shaped zigzag core and elongated alkyl chains.
The realization of the bandlike transport of \tm{26} has been confirmed unambiguously by observations of an ideal Hall effect, a negative temperature coefficient for mobility, and the Elliott--Yafet spin-relaxation mechanism~\cite{mitsui2014high, tsurumi2017coexistence}.
In addition, the transport characteristics, particularly in the single-crystal organic field-effect transistor (OFET) under systematic changes in its crystal structure, allow in-depth investigation of the intermolecular vibration because the intermolecular interaction in inherently soft OSCs can be tuned effectively by an external pressure or strain.
In fact, the mobility of the single-crystalline \tm{26} reaches \SI{16}{cm^2.V^{-1}.s^{-1}} and improves by a factor of 70\% with an $\sim$10\% suppression of intermolecular vibration at 3\% uniaxial strain~\cite{kubo2016suppressing}.
At low temperatures, $\sim$\SI{4}{K}, the mobility of the single-crystalline \tm{26} is expected to increase by 40 times that at room temperature and may reach \SI{650}{cm^2.V^{-1}.s^{-1}} by approximately 90\% suppression of the molecular vibration~\cite{tsurumi2017coexistence}.

Although intermolecular vibration concomitant with the thermal fluctuation of the transfer integral in OSCs has been intensively studied, methodologies for assessing the strength of intermolecular vibration have been limited so far. 
This is clearly because such intermolecular vibration modes in OSCs are likely to exist in terahertz regimes, at which the phonon energy is $\sim$\SI{10}{meV}.
There are several preliminary reports on the observation of the low-frequency intermolecular vibration modes using thermal diffuse electron scattering~\cite{eggeman2013measurement, illig2016reducing} and X-ray diffraction (XRD) in conjunction with translation--libration--screw analysis~\cite{schomaker1968rigid}.
However, these techniques are not always suitable for OSCs because these methods are performed under the assumption that molecules are rigid. This assumption is clearly unrealistic because most of the high-mobility OSCs recently developed possess a long alkyl chain, whose thermal motion should be independent of the $\pi$-core motions~\cite{illig2016reducing}.
Neutron scattering is also considered to be a promising technique for directly observing phonon modes in OSCs~\cite{harrelson2019direct}.
However, neutron scattering measurements require a large amount of material and have so far been limited to use with an OSC powder, which prevents an in-depth understanding of dynamic disorder under the application of strain.

Terahertz Raman spectroscopy enables the phonon energy at the $\Gamma$ point to be obtained because in principle the lattice vibration is described by phonon dispersion, and the momentum conservation law for phonon and photon should be satisfied before and after their inelastic scattering.
The phonon energy shift reflects the change in restoring force of each mode; thus, it is possible to track the shift in vibration amplitude for each Raman-active mode.
This feature is essential for the investigation of phonon scattering because a vibration mode with a large real-space amplitude of molecules does not necessarily correspond to a high degree of electron--phonon (e--ph) coupling.
Additionally, Raman spectroscopy can be carried out under ambient conditions with a small amount of material, which enables the measurement of single crystals.
Terahertz Raman spectroscopy has been performed on OSCs to investigate intermolecular vibrations~\cite{weinberg2007evidence, della2004intramolecular, sosorev2018relationship}.
However, Raman spectroscopy under strain has not yet been reported with OSCs, even though Raman spectroscopy is known to be a powerful tool for investigating the effect of strain on inorganic semiconductors~\cite{anastassakis1970effect, ni2008uniaxial}.
This is because a uniform strain cannot be applied to the conventional polycrystalline samples.

In this study, we performed terahertz Raman spectroscopy and molecular dynamics (MD) simulations for the state-of-the-art single-crystalline OSC \tm{26}.
The terahertz Raman modes were successfully observed down to \SI{10}{cm^{-1}} and were assigned to translational or rotational motions of the rigid $\pi$ core. 
Importantly, these terahertz modes were found to be amplified monotonically by uniaxial tensile strain, which demonstrates that the intermolecular vibration that particularly correlates with electron--phonon coupling can be tuned by adjusting the external strain. 

We first present how low-frequency Raman spectroscopy is sensitive to intermolecular vibration.
\tm{26} has two polymorphs (Supplemental Material (SM) Section A)~\cite{sm}; one is termed the ``lying'' structure, which is stable at temperatures below \SI{150}{K}, and the other is termed the ``standing'' structure, which is stable at temperatures above \SI{350}{K}. 
Note that these polymorphs are known to have the same space group P2$_1$/c according to single-crystal XRD measurements.
Although both polymorphs coexist at room temperature, it is possible to grow each single crystal separately by optimizing a recrystallization process.
The mobility for the standing polymorph reaches \SI{16}{cm^{2}.V^{-1}.s^{-1}}, whereas that for the lying structure is predicted to be \SI{1.5}{cm^{2}.V^{-1}.s^{-1}}~\cite{ishii2018quantitative}.
Terahertz Raman spectra were obtained for a bulk crystal [Fig.~\ref{fig:1}(b)] of the two polymorphs on a glass substrate (SM Section B).
There is a notable difference in the peak positions between the two spectra [Fig.~\ref{fig:1}(c)].
As the two polymorphs have the same symmetry, the difference in the peak positions is due to the difference in intermolecular interactions.
The observed lowest-energy peaks lie at \SI{9.7}{cm^{-1}} for the lying structure and at \SI{11.7}{cm^{-1}} for the standing structure.
These are in good agreement with the peaks obtained theoretically by MD simulation CONFLEX (CONFLEX Corporation), which predicted the positions of the lowest-energy peaks to be \SI{14.64}{cm^{-1}} for the lying structure and \SI{10.86}{cm^{-1}} for the standing structure.
Therefore, we conclude that all the Raman-active modes were successfully measured.
This enables the difference in intramolecular vibration to be distinguished with precision.
In the lying structure, the number of peaks below \SI{10}{meV} is significantly suppressed where the energy of low-frequency modes is sufficiently smaller than the thermal energy of room temperature, $\sim$\SI{25}{meV}. 
This suggests that these optical phonons are excited at room temperature, and the mobility of the lying structure may be limited by the phonon scattering because the intense peaks in the Raman spectra are likely to indicate strong of e--ph couplings~\cite{sosorev2018relationship}.

According to the MD simulation, the phonon modes can be classified into three categories: rigid $\pi$-core motions, $\pi$-core deformations, and covalent bond vibrations  (SM Section C).
Rigid $\pi$-core motions are observed below \SI{60}{cm^{-1}} [SM Fig.~3(a)].
The translational motion in the longitudinal direction lies at approximately \SI{20}{cm^{-1}}; here, the translational motions of molecules in a unit cell are opposite in phase. 
In addition, there are some translational modes in which the translational motions of the two molecules are in phase and the displacement of the alkyl chains is opposite to that of the $\pi$ cores to maintain the center of mass in the unit cell. 
Note that these do not involve a distinct intramolecular deformation.
$\pi$-core deformation, on the other hand, appears mainly above \SI{60}{cm^{-1}}; here, the vibration mode is assigned to a flat-spring-like bending deformation [SM Fig.~3(b)].
As the wavenumber increases further, the number of nodes of bending or stretching motion increases.
Thus, the motion of the higher-frequency modes is assigned to a vibration of the covalent bond above \SI{1000}{cm^{-1}}.

\begin{figure}
	\centering
	\includegraphics[pagebox=artbox]{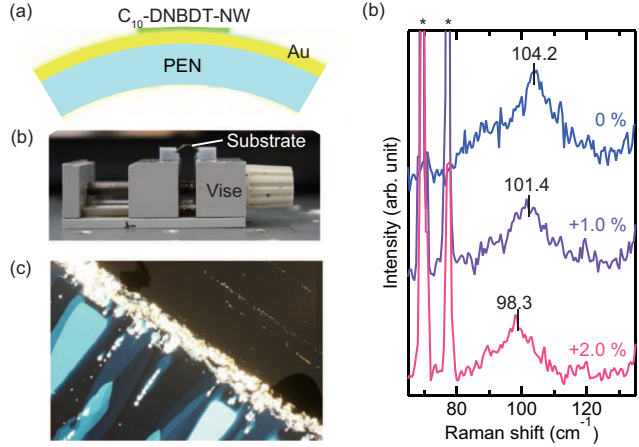}
	\caption{Terahertz Raman measurement under uniaxial strain. (a) Schematic of the sample. (b) Photograph of vise that supports the strained sample. (c) Cross-polarized microscopy image of single-crystalline thin film of \tm{26} fabricated by edge-casting method. (d) Terahertz Raman spectra of \tm{26} under three values of uniaxial strain.}\label{fig:2}
\end{figure}

To assess intermolecular vibration modes under uniaxial strain, terahertz Raman spectroscopy was also conducted for a single-crystalline thin film of \tm{26} on a flexible substrate (SM Section B).
First, a polyethylene naphthalate (PEN) substrate was coated with gold to prevent the Raman signal reaching from the PEN substrate.
The single-crystalline thin film of \tm{26} was then grown directly on the surface of the gold/PEN substrate by the edge-casting method~\cite{uemura2012band} [Fig.~\ref{fig:2}(a)].
A tensile uniaxial strain was applied to the crystal by bending the substrate as shown in Fig.~\ref{fig:2}(b).
The direction of the strain is parallel to the $c$-axis, as in our previous study~\cite{kubo2016suppressing}, in which the mobility was additionally evaluated along the $c$-axis as a function of the uniaxial strain.
In the present study, we applied only the tensile strain because the working distance of the objective lens is too short to allow compressive strain to be applied.

The spectra obtained for the various strains are shown in Fig.~\ref{fig:2}(d).
In this measurement, the lowest-energy peak, at \SI{11.7}{cm^{-1}}, is not resolved because the relatively weak Raman signals are overlapped by the background baseline due to the surface plasmon of gold.
The peak at \SI{104.2}{cm^{-1}} shows a clear red shift to \SI{98.3}{cm^{-1}} at 2.0\% tensile strain.
This indicates that intermolecular interaction and the amplitude of the intermolecular vibration modes can be tuned by applying external strain.

\begin{figure}
	\centering
	\includegraphics[pagebox=artbox]{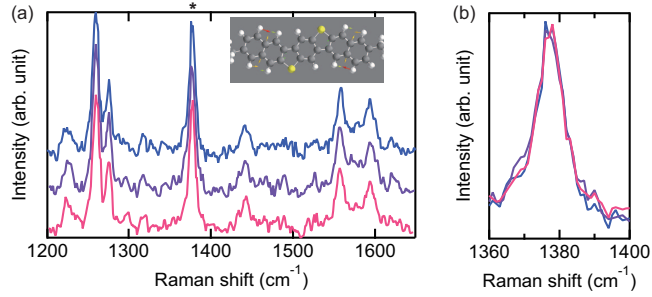}
	\caption{Strain dependence of Raman spectra in the high-frequency region. (a) Raman spectra in the high-frequency region, whose peaks correspond to intramolecular vibration modes. For example, the peak at \SI{1380}{cm^{-1}} (marked *) is assigned to the intramolecular vibration shown in the inset, in which arrows represent the displacement vector. The displacement vector of the intramolecular vibration modes was calculated by density functional theory using Gaussian~09 (Gaussian, Inc.) at the B3LYP/6-31(G) level. (b) Magnified Raman spectral peak marked as ``*'' as a function of uniaxial tensile strain.}\label{fig:3}
\end{figure}

However, no peak shifts are observed at high-frequency regimes, above \SI{1000}{cm^{-1}} [Figs.~\ref{fig:3}(a), (b)].
These high-frequency modes are assigned to the vibrational motion of the covalent bond, \textit{i.e.}, intramolecular vibration modes, which indicates that the strength of the covalent bonds is independent of uniaxial strain.

\begin{figure}
	\centering
	\includegraphics[pagebox=artbox]{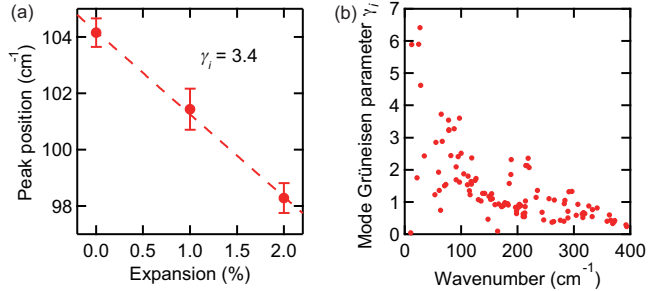}
	\caption{(a) Position of the terahertz Raman mode peak (\SI{104.2}{cm^{-1}}) as a function of uniaxial strain. The red line represents the linear fit, from which the Gr\"{u}neisen parameter was evaluated. (b) Simulated mode Gr\"{u}neisen parameters as a function of ambient wavenumber.}\label{fig:4}
\end{figure}

Here, we quantify the magnitude of the peak shift by calculating mode Gr\"{u}neisen parameter $\gamma_i$ for the peak at \SI{104.2}{cm^{-1}}.
The mode Gr\"{u}neisen parameter describes the anharmonicity of each phonon mode that causes coupling between the lattice volume and the eigenfrequency of the phonon.
The Gr\"{u}neisen parameter of a mode is defined as
\begin{align}
\frac{\Delta \omega_i}{\omega_i} &= - \gamma_i \frac{\Delta V}{V}\\
&= -\gamma_i (1-\nu) \epsilon,
\end{align}
where $i$ is the index of the mode, $\omega_i$ is the peak position at ambient strain, $\Delta \omega_i$ is the change in the peak position under the applied strain, $\Delta V / V$ is the proportional change in the sample volume, $\nu$ is the Poisson ratio, and $\epsilon$ is the applied strain (SM Section D).
According to the XRD measurement, the Poisson ratio $\nu$ of \tm{26} was estimated to be $\sim0.15$~\cite{kubo2016suppressing}, and thus the mode Gr\"{u}neisen parameter $\gamma_i$ is estimated to be $3.4$ from the line fitting of the Raman shift under the uniaxial strain (Fig.~\ref{fig:4}).

For many inorganic materials, particularly those whose crystals have only one type of interatom bonding, $\gamma_i$ is nearly constant for all modes.
For example, $\gamma_i$ for silicon and germanium was reported to be $\sim1$~\cite{mitra1969pressure, bhattacharyya2010anomalous}, and the value for GaAs and ZnO is approximately 1.5~\cite{bhattacharyya2010anomalous,adachi1994gaas}.
Aside from semiconductors, a $\gamma_i$ value of $3.0$ was reported for the ionic crystal NaF~\cite{mitra1969pressure}.
In contrast, $\gamma_i$ of OSCs is mode-to-mode dependent because hard intramolecular vibrations due to covalent bonding and soft intermolecular vibrations due to van der Waals interactions coexist in OSCs.
For these, a relatively high $\gamma_i$ value of $>3$ below \SI{30}{cm^{-1}} was observed in molecular crystals such as S$_8$ and As$_2$S$_3$~\cite{1974zallenpressure}.
The length of soft bonds can change more dramatically than that of hard bonds when external strain is applied.
Therefore, weak intramolecular interactions can be effectively tuned using the pressure applied by strain.

In order to confirm the mode-to-mode dependence of $\gamma_i$, MD simulation was carried out using CONFLEX\@.
All the optical modes at the $\Gamma$ point, including Raman-inactive modes, were calculated after performing the intra-lattice structure optimization under the fixed lattice constant (SM Section C).
The simulated $\gamma_i$ shows considerable mode-to-mode dependence [Fig.~\ref{fig:4}(b)].
The mode Gr\"{u}neisen parameter at \SI{104.2}{cm^{-1}} is estimated to be approximately 3.3, which is in good agreement with the value experimentally obtained, $3.4$.
However, the calculated $\gamma_i$ shows large dispersion, which is in contrast to the monotonic frequency dependence observed with S$_8$ and As$_2$S$_3$ under hydrostatic pressure~\cite{1974zallenpressure}.
The large dispersion of $\gamma_i$ arises because we applied uniaxial strain instead of hydrostatic pressure.
The modes with the large $\gamma_i$ values are likely to have the large displacement of $\pi$-core/alkyl chains parallel to the strain.
On the other hand, the lowest-energy mode shows the translational vibration of $\pi$ cores perpendicular to the strain [SM Fig.~3(a)], whose $\gamma_i$ is close to zero because the intramolecular interaction is less modified perpendicular to the strain.

Here, we compare the energy shift of low-frequency phonon modes with the change in mobility with respect to the uniaxial strain.
The strain dependence of mobility showed linear behavior near $\epsilon=0$ in our previous study~\cite{kubo2016suppressing}, and the change $g$ in the mobility ($g=\mu_0^{-1}\left.\partial \mu/ \partial \epsilon\right|_{\epsilon=0}$) was estimated to be $13.3$, where $\mu_0$ and $\mu$ are the mobility at ambient strain and that under the applied strain, respectively.
If the mobility of OSCs is determined by the phonon scattering's having a particular dominant vibration mode $i$, the temperature dependence of mobility follows the power law $\mu \propto T^{\alpha}$.
This assumption gives $g$ as the following equation (SM Section E):
\begin{align}
g=\frac{1}{\mu_0}\left.\frac{\partial\mu}{\partial \epsilon}\right|_{\epsilon=0}  &= 2 \alpha \gamma_i (1-\nu).
\end{align}
Using the experimentally obtained mode Gr\"{u}neisen parameter $\gamma_i$ value of $3.4$ and the exponent $\alpha$ value of $-0.85$ for the temperature dependence of mobility from our previous paper~\cite{tsurumi2017coexistence}, the mobility change $g$ under the uniaxial strain is estimated to be $4.8$, which is considerably smaller than the experimentally obtained value, $13.3$.
This simple model calculation indicates that rather than a single vibrational mode, multiple modes should be concomitant in the phonon scattering.
Indeed, there are several vibration modes with $\gamma_i$ greater than 6, which agrees fairly well with experimental observation, though not perfectly.
For example, when $\gamma_i=6.4$ is used, $g$ is estimated to be $9.3$, which is not in perfect agreement with the experimentally observed value.
This discrepancy may originate from the assumption that momentum relaxation is limited only by a single intermolecular vibration mode.
It may be more reasonable to assume that multiple vibration modes have an impact on carrier relaxation.
Overall experimental and calculational results reveal that one of the translational or rotational modes that appear in the terahertz frequency range plays a vital role in limiting the mobility.

In summary, we performed terahertz Raman spectroscopy on single-crystal \tm{26} and successfully observed the terahertz Raman modes down to \SI{10}{cm^{-1}}.
The observed intermolecular vibration mode showed a clear red shift under uniaxial tensile strain, and a large mode Gr\"{u}neisen parameter $\gamma_i$ value of $3.4$ was obtained, indicating that intermolecular interaction and the amplitude of the molecular fluctuation can be tuned by the application of external strain.
The MD simulation revealed that the molecular fluctuation as modified under uniaxial strain is qualitatively consistent with the effect of strain on mobility.
We expect that terahertz Raman spectroscopy under uniaxial strain will become a powerful tool for the further investigation of structure--property relationships of OSCs when accurate peak assignment becomes possible.

This work was supported by JSPS KAKENHI grants (nos. JP18H05954, JP19K15652,  JP17H06123, JP17H06200, and JP17H03104). S.W. acknowledges support from the Leading Initiative for Excellent Young Researchers of JSPS\@.

\bibliographystyle{apsrev4-1}
\bibliography{references.bib}
\end{document}

%% file: settings.tex
\usepackage{amsmath}
\usepackage{siunitx}
\usepackage{graphicx}

\newcommand{\tm}[1]{%
	\ifnum #1=26 C$_{\text{10}}$--DNBDT--NW%
}

\newcommand{\tmfull}[1]{%
	\ifnum #1=26 3,\hspace{0pt}11-\hspace{0pt}didecyl\hspace{0pt}dinaphtho\hspace{0pt}[2,\hspace{0pt}3-\hspace{0pt}\textit{d}:\hspace{0pt}2',\hspace{0pt}3'-\hspace{0pt}\textit{d}']\hspace{0pt}benzo\hspace{0pt}[1,\hspace{0pt}2-\hspace{0pt}\textit{b}:\hspace{0pt}4,\hspace{0pt}5-\textit{b}']\hspace{0pt}dithiophene%
}